# The Red Day Star, the Women's Star and Venus: D(L/N)akota, Ojibwe and Other Indigenous Star Knowledge

Annette S. Lee, St. Cloud State University, USA
Jim Rock, Augsburg College, USA
William Wilson, St. Cloud State University, Fond du Lac Tribal & Community College, USA
Carl Gawboy, St. Cloud State University, Fond du Lac Tribal & Community College, USA

Abstract: In Ojibwe the Morning Star is called *I'kwe Anung*, which means the Women's Star. In D(L/N)akota the same planet Venus is called *Aŋpetu D/Luta Wiçaŋlipi* the Red Day Star. Both cultures have rich and interesting understandings of Venus that relate to other Indigenous cultures throughout the world. Venus is so often related to the feminine because native peoples carefully watched the movement of the 'star' and saw it in the east at sunrise for nine months and then in the west at sunset for the following nine months. Nine months is exactly the time for human gestation. Yet, tragically, the native star knowledge is disappearing as elders pass. The Native Skywatchers project focuses on understanding the *Ojibwe* and D(L/N)akota importance of this and other celestial connections. MN State Science Standards K-12 requires "Understanding that men and women throughout the history of all cultures, including Minnesota American Indian tribes and communities, have been involved in engineering design and scientific inquiry….For example *Ojibwe* and Dakota knowledge and use of patterns in the stars to predict and plan." And yet there is a complete lack of materials. Working closely with a team of culture teachers and language experts we are building community around the native star knowledge.

## Introduction and Purpose

In Ojibwe the Big Dipper is *Ojiig* – the Fisher (Morton and Gawboy 2000, 175-177; Gawboy 2005) and in D(L/N)akota star knowledge the same group of stars is seen as *To Wiŋ/Toŋ Wiŋ*—Blue Woman/Birth Woman (Goodman 1992, 22). In each there are stories and teachings that help guide, teach and inspire native peoples. The Native Skywatchers Project focuses on understanding the Ojibwe and D(L/N)akota importance of these and other celestial connections. We seek to address the crisis of the loss of the indigenous star knowledge, specifically the native peoples of Minnesota, Dakota and Ojibwe. The purpose of this programming is to remember, rebuild and revitalize the native star knowledge.

There is urgency to this project for two reasons: the native star knowledge is disappearing as elders pass and state standards. One Ojibwe elder spoke of his vision of 'the star medicine returning through the native youth.' He specifically called them 'star readers' (P. Schultz, pers. comm.). In 2011 he passed away suddenly. At the same time, the new MN State Science Standards K-12 requires "Understanding that men and women throughout the history of all cultures, including Minnesota American Indian tribes and communities, have been involved in engineering design and scientific inquiry….For example Ojibwe and Dakota knowledge and use of patterns in the stars to predict and plan" (Minnesota Department of Education, 2010). And yet there is a complete lack of materials.

This research, The Native Skywatchers Project, seeks out elders, culture teachers and language experts to discuss the Ojibwe and D(L/N)akota star knowledge. From these sources and working with the elders we have created two astronomically accurate, culturally important star maps, *Ojibwe Giizhig Anung Masinaaigan* – Ojibwe Sky Star Map and *Makoçe Wiçaŋlipi Wowapi* – D(L)akota Sky Star Map. These valuable maps were disseminated to regional

The International Journal of Science in Society
Volume 4, 2013, www.science-society.com, ISSN 1836-6236
© Common Ground, Annette S. Lee, Jim Rock, William Wilson, and Carl Gawboy
All Rights Reserved, Permissions: cg-support@commongroundpublishing.com

educators at the *Native Skywatchers Middle School Teacher* workshop June 2012. In addition, hands-on curriculum that combines astronomy, culture, language and art has been developed. As with many North American tribes much cultural knowledge, especially cultural astronomy, has been lost. The goal of the *Native Skywatchers* programming is to build community around the native star knowledge.

## Procedures and Methods

The *Native Skywatchers* programming is led by the author, Annette Lee, Professor of Astronomy and Planetarium Director at St. Cloud State University. Funding has been provided by: NASA-MN Space Grant, North Star STEM Alliance, St. Cloud State University and Fond du Lac Tribal and Community College. This project represents a unique collaboration between a large state university, a tribal and community college and federal agencies. The strategy of the project is to combine astronomical expertise, cultural knowledge and artistic talents to create programming such as star maps, constellation guides, and curriculum that currently does not exist. Regional teachers requesting information relating to the native star knowledge also motivated the creation of the map and related curriculum. This was clearly related to the new MN State Science Standards K-12, in particular benchmark 3.1.3.2.1 *"Understand that everybody can use evidence to learn about the natural world, identify patterns in nature, and develop tools"* (Minnesota Department of Education, 2010).

Travel and interviews were conducted to consult with various Ojibwe and D(L/N)akota cultural experts, such as Carl Gawboy (Boise Forte), Paul Schultz (White Earth), and William Wilson (Lake Nipigon), Jim Rock (Dakota), Charlene O'Rourke (Pine Ridge), Albert White Hat (Rosebud), Duane Hollow Horn Bear (Rosebud) and Arvol Lookinghorse (Green Grass) over a three-year period. Astronomical, language and cultural teachings were shared and recorded.

A culminating focus of the Native Skywatchers project was the creation of two star maps: *Ojibwe Giizhig Anung Masinaaigan* – Ojibwe Sky Star Map and *Makoċe Wiċaŋḣpi Wowapi – D(L/N)akota Sky Star Map.* The maps are organized with Polaris –the North Star in the center. This is to emphasize the closeness of Polaris to our current north celestial pole (NCP) and circumpolar motion. Because of circumpolar motion, we appear to see all the stars in the night sky revolve around the North Star in a counter-clock wise motion as the hours pass each night into day. Because of this motion, in some native cultures the North Star is seen as one of the leaders of the star nation. The 'Northern Stars' referred to here by the Ojibwe and D(L/N)akota people are the circumpolar stars as seen from approximately 45-55º N latitude, 85-110º W longitude.

All stars not circumpolar, as seen from 45-55º N latitude, will rise in the east and set in the west at regular times throughout the year. They are seasonal stars. The *Ojibwe Giizhig Anung Masinaigan* – Ojibwe Sky Star Map and *Makoċe Wiċaŋḣpi Wowapi – D(L)akota Sky Star Map* are arranged in order to show the constellations that are best visible each season. This assumes a viewing time of about two to three hours after sunset. In the night sky stars of each season can be best seen overhead or in the south during that particular season. For example if you look at the stars in the early summertime, a few hours after sunset you will see Hercules overhead and Scorpio low on the southern horizon. These are early summer stars.

## Results

### *Venus*

Venus is the third brightest object in the sky after the Sun and the Moon. It is so bright that it is often mistaken for a UFO. This is because Venus is the closest planet to us at only about 25 million miles (compared to Mars at 47 million miles at closest approach). Its brightness is completely due to light reflected from the Sun, like Earth and all terrestrial planets Venus does not generate any visible light. Venus' light is especially bright because it is covered in white, reflective clouds. For this reason, people had a fascination with 'life on Venus' in the early twentieth century, as we could not see what was underneath the clouds. In the 1960's the first missions used radar wavelengths to cut through the cloud layer and see the topography underneath. We found extensive volcanism (Head, J. et al. 1992).

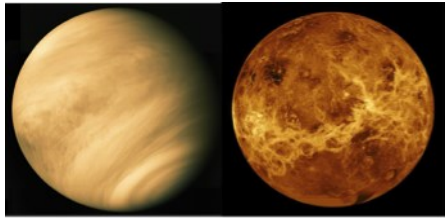

Fig. 1: Venus Visible wavelengths (left), Radar (right) (Photo courtesy of NASA)

## Ojibwe Connection to Venus

In Ojibwe star knowledge, Venus is known as *I'kwe Anung* – Women's Star (C. Gawboy, unpublished data). This name has multiple layers of meaning. The first understanding is that native Ojibwe people carefully observed the motion of Venus each day/night and found patterns in the movement. The pattern of Venus' movement as seen from an observer on Earth is that Venus will appear in the east before sunrise (the Morning Star) and then in the west just after sunset (the Evening Star). The pattern repeats on a nine-month cycle. A person can watch Venus in the morning for about nine months; it disappears for a short time and then reappears in the opposite sky at sunset for about nine months. Then the cycle repeats. Remarkably this nine-month cycle is the same time as for human gestation. This is why Ojibwe and other indigenous cultures have associated Venus with the feminine.

Another connection between Venus and women is that traditional Ojibwe women were responsible for gathering the water. Very commonly women would rise very early before sunrise to gather water for the camp. (C. Gawboy, pers. comm.) It should also be noted that Venus is also called *Waabun'Anung* – East Star, which is translated 'Morning Star'. This is due to the fact that Venus is visible for about half of its cycle in the east just before sunrise.

## D(L/N)akota Connection to Venus

In D(L/N)akota, the word for Venus is *Aŋpetu D/Luta Wiçaŋhpi*, which translates 'Red Day Star' (A. Lookinghorse, pers. comm.). This idea has several layers. The first understanding is literal. The 'star-like' point of light that is the planet Venus, and the surrounding sky appear reddish in color. This is because Venus can only be seen low on the east or west horizon at sunrise or sunset. When celestial objects are positioned low in the sky (or just above the horizon), they appear reddish. This is due to Rayleigh scattering. The longer path of sunlight through the atmosphere at low elevations removes almost all of the blue and green light.

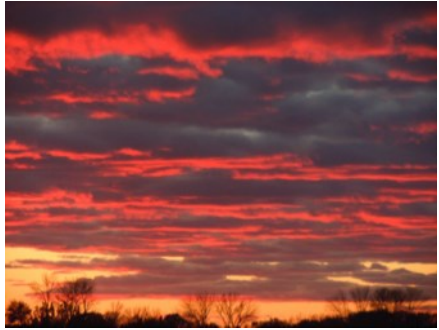

Fig. 2: Scattering of Sunlight at Sunset. (Photo by A. Lee.)

Another layer of meaning in *Aŋpetu D/Luta Wiçaŋḣpi* – Red Day Star requires a cultural context. Red is a sacred color, to say a 'red day' is equivalent to ' a sacred day'. For example, a traditional D(L/N)akota person wakes up gives thanks for life and the chance to live another day (A. Lookinghorse, pers. comm.). Recently, people refer to walking on the 'red road'. It is a way of remembering our connection with all living beings, also stated *mitakuye oyasin/owasiŋ*—all my relatives. *Aŋpetu D/Luta Wiçaŋḣpi* – Red Day Star is considered one of the leaders of the star nation (A. Lookinghorse, pers. comm.). All families with the name *D/Luta* – Red were keepers of the star knowledge, for example: Red Eagle, Red Deer, Red Willow, Red Horse, Red Day and so on. These families traditionally protected and passed down the star knowledge from generation to generation (A. Lookinghorse, pers. comm., J. Rock, pers. comm.). Furthermore traditional star quilts were made with one large star in the center, this was in honor of *Aŋpetu D/Luta Wiçaŋḣpi* – Red Day Star (A. Lookinghorse, pers. comm.).

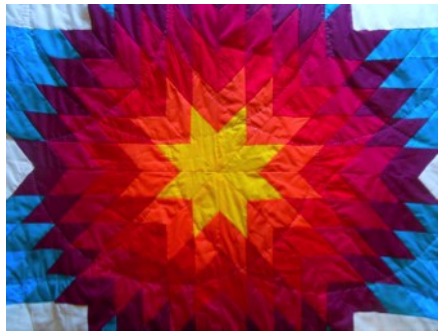

Fig. 3: Star Quilt. (Photo by A. Lee.)

Much of the D(L/N)akota star knowledge rests on the framework of mirroring, which is said, "As it is above (in the sky), it is below (on the Earth)." *Kapemni* – can mean swinging around or mirroring in D(L/N)akota, which is drawn by two tipis/triangles connected at their apexes. When D(L/N)akota people on Earth mirror what is happening in the stars, a spiritual doorway opens up because of this connection. It is understood that the healing power of the star nation, '*the woniya of Wakaŋ Taŋka*', flows through (Goodman 1992, 31-34).

Another word for morning star is *Aŋpo Wiçaŋḣpi* or *Aŋpao Wiçaŋḣpi* (J. Rock, pers. comm.). This refers to dawn or 'as the morning comes' (Buechel and Manhart 2002, 21). In this context *Aŋpo Wiçaŋḣpi* specifically refers to Venus as the Morning Star as opposed to *Wiçaŋḣpi Haŋyetu* or *Haŋyetu Wiçaŋḣpi* as Evening Star. Interestingly the star Arcturus in the Greek constellation Bootes is called *Aŋpo Suŋkaku Wiçaŋlipi*, which means Younger Brother of Morning Star (J. Rock, pers. comm.).

## Other Cultural Connections to Venus

Various cultures throughout the world and spanning history have related connections to Venus. The Maya had entire codices devoted to the observational movements of Venus. They kept track of every astronomical event related to Venus: inferior conjunctions, superior conjunctions, transits, synodic orbit, etc. There are many Mayan words for Venus such as: *Noh Ek* (Big Star), *Chak Ek* (Red or Great Star), *Sastal Ek* (Bright Star) and *Xux Ek* (Wasp Star). Venus is associated with abundance, fertility, growth, death and rebirth (Milbrath 1999, 34-36). Dakota astronomer, Jim Rock, also associates Venus with related *Çekpa* Dakota and Maya traditions. (Rock 1997).

Another example of indigenous star knowledge relating to Venus comes from the Aboriginal Australian Yolngu people who know Venus as *Banumbirr*. They conduct a 'Morning Star Ceremony' that serves to help relatives communicate with those passed away. The ceremony goes from sunset to sunrise when Venus first re-appears as the Morning Star. On a really dark, clear night Venus can be seen with a faint band of light that stretches down towards Earth, zodiacal light. This is understood as *Banumbirr* and a rope that connects her to the island of Baralku. There is a strong component of mirroring in the ceremony with a Morning Star Pole that includes Venus and the rope (Norris 2009, 18-22).

Lastly, the Greeks associated the planet Venus with the goddess of love, Aphrodite. Later the Romans changed the planet's name to the goddess Venus (Krupp 1992, 177). Both of these have a strong, clear connection to the feminine. Venus is the only planet in our Solar System with a feminine name. In addition, all but three of the surface features on Venus were given feminine names as declared in 1919 by the International Astronomical Union.

### *Ojibwe Star Map*

The Ojibwe map is a collaborative work between Annette Lee, William Wilson and Carl Gawboy. The Ojibwe constellations are a result of the research and work of Carl Gawboy over a forty-year period (C. Gawboy, unpublished data). Traditional Ojibwe x-ray style was used by William Wilson to paint the Ojibwe constellations. It is symbolic of seeing the unseen. It is an allegory for the material world and the spirit world. The brightly colored internal organs and anatomical shapes are a glimpse into the inner layers of our bodies. "We are seeing the picture as the spirits see us. They see right through. The strange looking animals and figures are portrayed as they come in ceremony. Sometimes they are half beaver, half eagle. Sometimes they are scary. Sometimes tempting.", explains William Wilson (W. Wilson, pers. comm.).

The border includes: strawberries, raspberries, blueberries and winterberries (traditional Ojibwe foods) illustrated in reference to Ojibwe style floral beadwork. Often the floral beadwork is done on black velvet or with a white beadwork background. Usually beadwork is done on items of importance spiritually or socially like pipe bags, moccasins, leggings, etc.

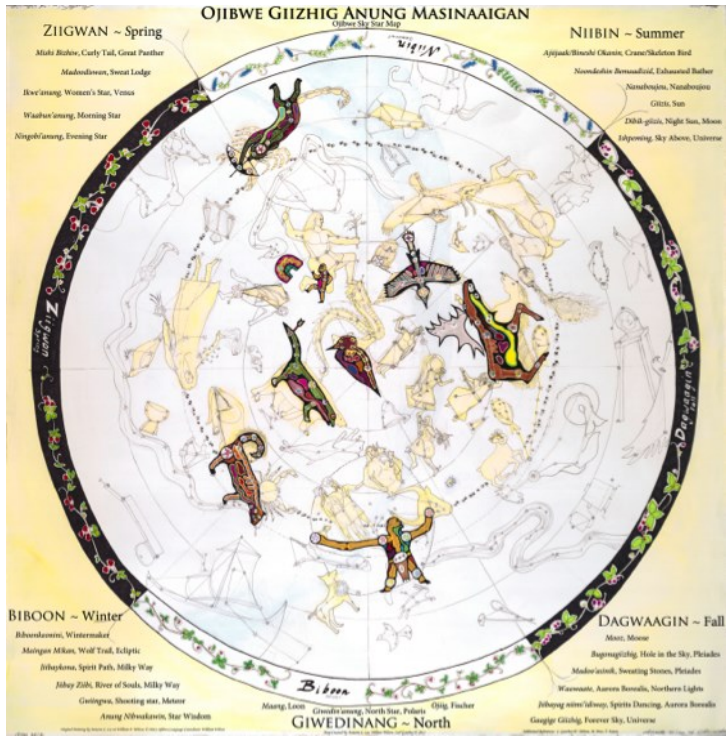

Fig. 4: Ojibwe Star Map created by A. Lee, W. Wilson, C. Gawboy (Photo by A. Lee.)

## Ojibwe Fall Stars

*Dagwaagin* is the Ojibwe word for fall. On the right quadrant of the map are the constellations best seen in the fall. The four bright stars in the shape of a square indicate the Ojibwe *Mooz* – Moose constellation. In Greek mythology these are the brightest stars of Pegasus, the winged horse. The moose is complete with legs stretching into Pisces, the moose's head pointing westward towards Cygnus, the swan and the moose's horns overlap with the Greek constellation, Lacerta. Located inside the large square are three stars that mark the moose's heart. This fall constellation can be seen in the Ojibwe pictographs at the Boundary Waters Canoe Area in northern Minnesota (Morton and Gawboy 2000, 189-191). High up on the cliffs at North Lake Hegman, the *Mooz* constellation is painted on the rock face, complete with heart line of stars indicated.

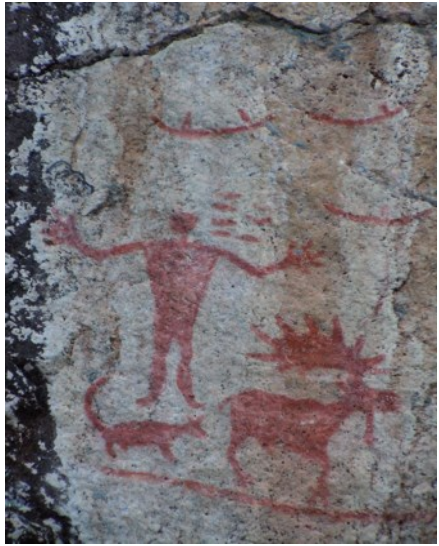

Fig. 5: Lake Hegman, Ojibwe Pictographs at Lake Hegman (Photo by A. Lee.)

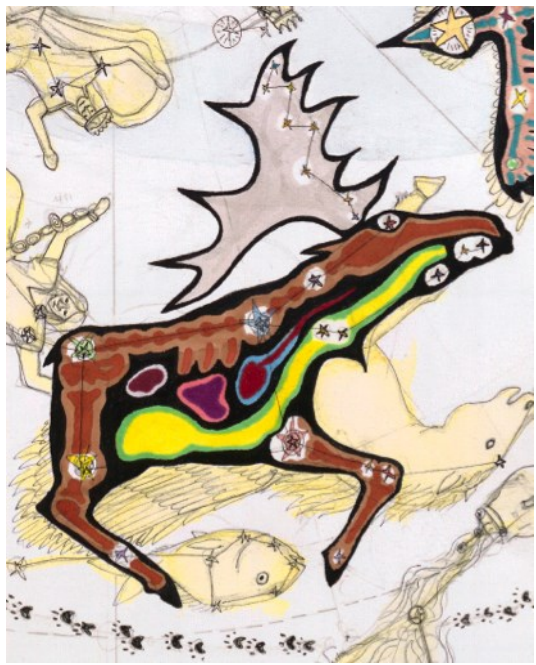

Fig. 6: *Mooz* – Moose Constellation (Photo by A. Lee.)

## Ojibwe North Stars

*Giwedin* is the Ojibwe word for North. These important stars are found in the center of the map. The Ojibwe fisher, *Ojiig* is found overlaying the Greek constellation Ursa Major (which includes the asterism Big Dipper). Here is an example of the keen observation skills of the Ojibwe people. The behavior of the fisher mirrors the motion of the stars. The fisher is known to be neither nocturnal nor diurnal in its sleeping and hunting patterns (W. Wilson, pers. comm.). It hunts, sleeps, then hunts again, not returning to the same den; is it constantly on the move. Those that watch the stars know that most stars are seasonal, but only those found within 45-55º of the North Celestial Pole will be circumpolar. This set of stars is referred to here as 'the North Stars'. They appear to rotate counterclockwise about Polaris in a regular 24-hour year-round pattern. In addition, the fisher is a small but ferocious fighter known for its ability to kill porcupines. This relates to some traditional Ojibwe stories that involve the heroic acts of the fisher. The only star that appears not to move in the northern hemisphere night sky, Polaris, is part of the Ojibwe constellation, *Maang* or loon. The loon constellation relates to the Ojibwe clan system. Loons are the leaders of the people (Benton-Banai 1988, 74-78).

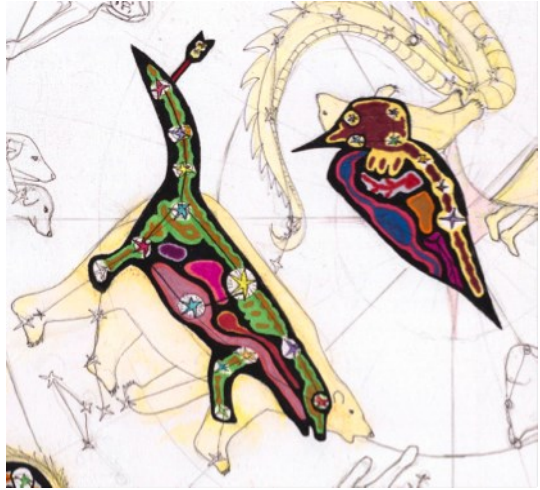

Fig. 7: *Ojiig* – Fisher and *Maang* – Loon constellations (Photo by A. Lee.)

## D(L/N)akota Star Map

The D(L/N)akota map, *Makoċe Wičaŋhpi Wowapi* was painted by the author, Annette Lee and the Dakota astronomy and language consultant was Jim Rock. The map is based on the chart found in the book "Lakota Star Knowledge" by R. Goodman and his interviews with many Lakota elders (Goodman 1992, 65). The star map was painted in reference to D(L/N)akota beadwork. It is said that each bead is a prayer. Beads are traditionally used to beautify sacred items like medicine bags and pipe stems. Also beadwork is used to adorn clothing or accessories, like on an outfit to wear to a special occasion. Beaded items are worn with great pride, for example, pow-wow regalia. The pinpoints of colorful dots in beadwork are reminiscent of starlight. The process of doing beadwork is meticulous and disciplined; it requires stillness. This stillness is echoed in the night sky. Beadwork and stars both sparkle.

The four directions are seen as spiritual and physical guideposts. Often seven directions are used which includes the four cardinal directions plus above, below and center. Many ceremonies and everyday prayers use the directions to focus and send the prayers. Albert White Hat explains the four directions as *Tateuye topa* - the four winds. The cardinal direction North in particular is associated with the wintertime and stillness. Trees and plants appear 'dead' on the outside in the winter, but they are still alive and growing on the inside, especially the roots. In wintertime people look to follow this example and practice stillness that will nurture inner growth. Each of the solstices and equinoxes mark the beginning of a season and are considered sacred days and good times to pray and have ceremony. Albert White Hat explains, "The seasons are described as births. Every season is a new birth" (White Hat Sr. 1999, 93-94).

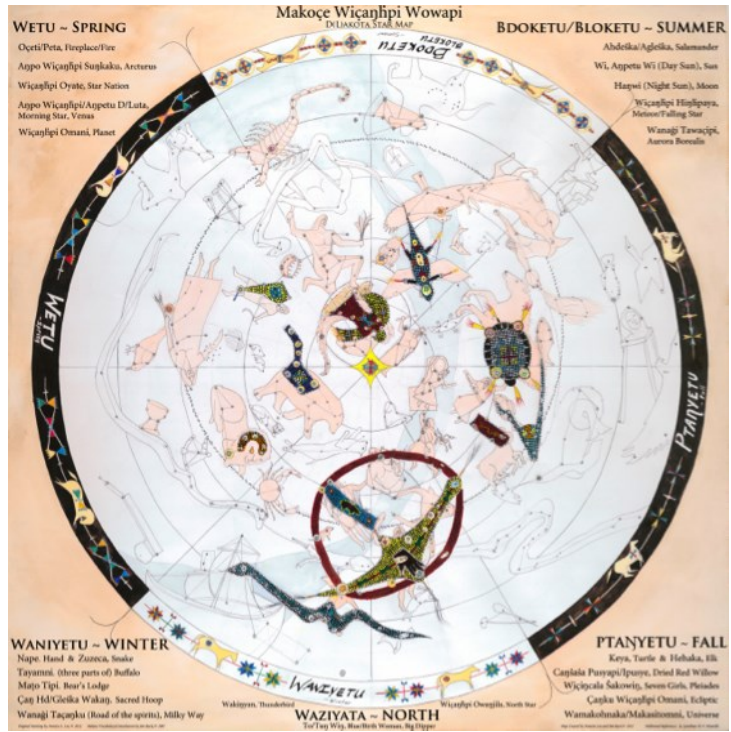

Fig. 8: D(L/N)akota Star Map created by A. Lee, J. Rock (Photo by A. Lee.)

**D(L/N)akota Fall Stars**

*Ptaŋyetu* is the D(L/N)akota word for fall. On the right quadrant of the map the D(L/N)akota constellations that are best seen in the fall are located as follows: *Keya* – Turtle, *Hehaka* – Elk and *Caŋśaśa Ipusye* – Dried Willow. The Turtle, *Keya* constellation refers to the medicine bag made for girls. The turtle's attributes of long life, steadfastness and fortitude are connected with the feminine. (Goodman 1992, 37-41) *Hehaka* – elk is associated with love and romance. (J. Rock, pers. comm.) *Caŋśaśa Ipusye* – Dried Willow (Red Osier Dogwood) is plant medicine used for the sacred pipe and other ceremonies.

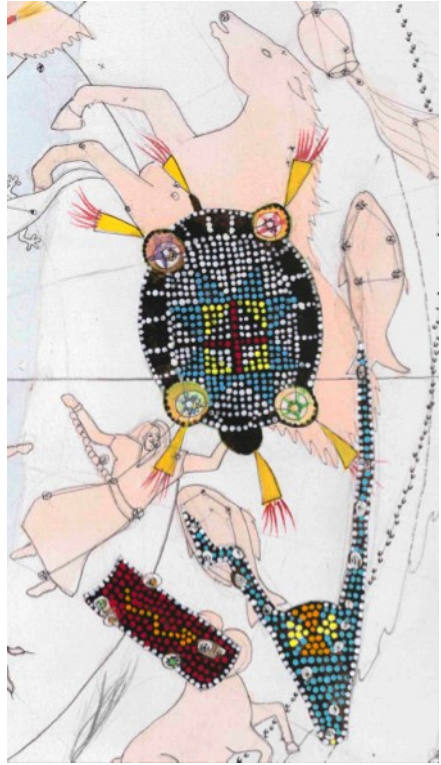

Fig. 9: *Keya* – Turtle, *Hehaka* – Elk and *Caŋśaśa Ipusye* – Dried Willow constellations (Photo by A. Lee.)

### D(L/N)akota North Stars

*Waziyata* is the D(L/N)akota word for North. Located in the center of the map are the circumpolar stars that are seen all seasons from the Northern Hemisphere. Over the north celestial pole is found *Wiçaŋḣpi Owaŋźila/Owaŋźina/Owaŋźida*, the North Star (the Star that Stands in One Place), *Wakiŋyaŋ* –Thunderbird, and *To Wiŋ/Tuŋ Wiŋ* - Blue Woman/Birth Woman overlaps with the Big Dipper/Ursa Major constellation. There is a depth of knowledge contained here in the northern night sky. Simply put, *To Wiŋ/Tuŋ Wiŋ* - Blue Woman/Birth Woman is a spirit that acts as a doorkeeper between the spirit world and the material world. She helps those crossing between the worlds; i.e. babies and recently deceased. The *Wakiŋyaŋ* – Thunderbird constellation overlaps with the Greek constellation Draco the dragon. *Wakiŋyaŋ* – Thunderbird constellation is also at the center of the precession circle. The Sun and Moon's gravitational pull on the Earth cause a small wobble in the Earth's axial tilt over a 26,000-year period. The heart of the Thunderbird, *Wakiŋyaŋ* constellation is at the center of this precession circle in the northern sky. This is an excellent example of the keen awareness and deep astronomical knowledge of the D(L/N)akota people. The word *Wakaŋ* means "the power to give life or to take it" (A. White Hat Sr., pers. comm.) and associating the thunder beings with both destructive and life giving powers is understood on several levels. On Earth, lightning and thunderstorms are often destructive forces causing forest and home fires. On the other hand, this same force can be a cleansing part of a natural cycle of growth, for example, prescribed burns. Astronomically there is another important connection, the stable axial tilt of the Earth over long periods of time is thought to be an important factor in the habitability of life on a planet, indeed some theories suggest that life itself might have been sparked in the primordial soup by lightning (Cairns-Smith et al. 1992, 161-180).

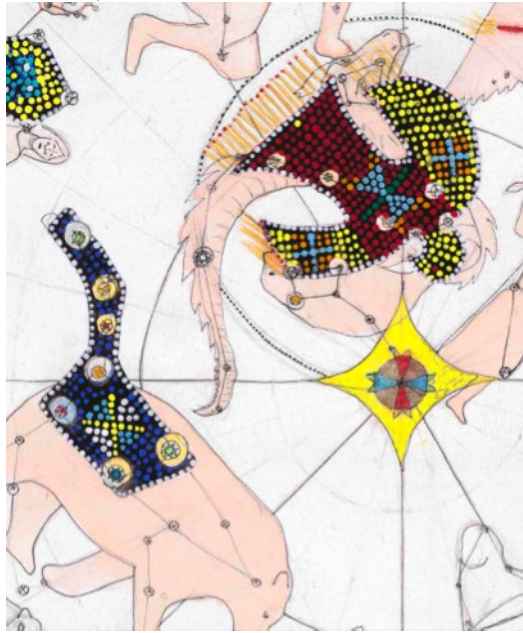

Fig. 10: *Wiçaŋḣpi Owaŋżila/Owaŋżina/Owaŋżida,* the North Star, *Wakiŋyaŋ* –Thunderbird, *To Wiŋ/Tuŋ Wiŋ* - Blue Woman/Birth Woman Constellations (Photo by A. Lee.)

## Conclusion and Future Directions

The methods presented here are interdisciplinary. Astronomy, culture, art and language are all represented in this research. And yet the delivery of such an in-depth, interdisciplinary topic like indigenous astronomy can be overwhelming to students, adults or youth, that have grown up with light pollution, tall buildings, and computers. Unlike traditional native people, current members of society spend a lot of time indoors. Most people have some familiarity with the Big Dipper, Sun and Moon. The delivery of this culturally rich material must be simple and yet allow for complexity and abstraction. To achieve this goal we first use the cultural framework of the four directions. The current night sky is subdivided into: north, east, south/overhead, and west. From the beginning of the discussion the cultural context is intact. The four directions are considered important framework and guideposts in native culture. This instructional approach builds on a sense of place that often native peoples are aware of (Semken 2005) and allows participants to connect the current night sky to sense of place. This technique grounds the complexity of the current night sky in the tangible and the simple, and yet allows for a multi-layered, circular learning approach. Following this approach allows for the widest range of participants to take part in the learning experience.

Furthermore, the stars and constellations can be best understood in terms of the four seasons. The discussion is simplified again by fixing the time as a few hours after sunset. This is the observing time, and is referred to as 'prime time' for stargazing. Only in the northern direction will the circumpolar stars, or North Stars as seen from approximately 45-55º N latitude, 85-110º W longitude, be visible throughout the year. When an observer faces due south (azimuth 180º along the horizon) he/she will see the current season of stars. The previous season will be seen setting in the west and the following season will be seen rising in the east. The *Ojibwe Giizhig Anung Masinaaigan* – Ojibwe Sky Star Map and the *Makoçe Wiçaŋḣpi Wowapi – D(L)akota Sky*

*Star Map* are best presented by transforming the discussion into an experiential, hands-on event. In addition, this highly visual, holistic and cooperative learning environment is more consistent with a traditional native learning style (Cleary and Peacock 1997).

Lastly, the *Native Skywatchers Project* is a collaborative approach. Native knowledge is sometimes a different way of knowing than Western science. There are strict cultural protocols that must be respected, such as some stories are to be told only when there is snow on the ground. We must be extremely careful not to introduce or propagate error into the written or oral records. Use caution and be hesitant. The orthography used here is 'D(L/N)' to represent the Siouan language spoken by the *Oceti Sakowin Oyate* – the People of the Seven Starfire/Campfire Nations - Dakota, Lakota, and Nakota nations. Also note that sometimes terms are now heard with the adjective preceding the noun, although traditionally the reverse was true. Users of these materials are urged to seek out elders and native community members to bring into the classroom. Materials represented here should be viewed as a beginning.

## Acknowledgements

The *Native Skywatchers* project acknowledges the elders and others that have kept this star knowledge alive. We acknowledge Paul Schultz (White Earth) who passed away suddenly in 2011 and Albert White Hat Sr. (Rosebud) who passed away in June 2013. Both men were collaborators with this project.

# ABOUT THE AUTHORS

*Annette S. Lee:* Annette Lee (mixed-race Dakota-Sioux) is an assistant Professor of Astronomy and Physics and Director of the Planetarium at St. Cloud State University in central Minnesota. The crisis my work addresses is preventing the loss of Ojibwe and Dakota/Lakota star knowledge. Elders are passing. Otherwise knowledgeable native elders tell me that they "…just weren't listening when the star stories were being told." Others talk of new generations of 'star readers' and how the star medicine will be brought back by the younger generation. Interwoven in the star knowledge is the language, which holds keen insight and observation far beyond what people practice today. My goal is to help preserve indigenous astronomy and pass it on to present and future generations. The Native Skywatchers Program is about creating sustainable change by building community around the native star knowledge. Having graduate degrees in Astrophysics (Washington Univ. 2009) and Painting (Yale 2000) and a UC-Berkeley alumni in Applied Mathematics (1992) the Native Skywatchers project bridges many worlds—learning from elders; relating Native star knowledge to Western knowledge; inspiring youth in science; engaging audiences through culture, art and science.

*Jim Rock:* Jim Rock (Dakota) has a Master's degree in education and has taught astronomy, chemistry and physics for thirty years for thousands of students in universities and high schools from urban, suburban and reservation communities. He currently teaches a Native Skywatchers course at Augsburg College offering indigenous cosmology lessons to teachers throughout Minnesota in collaboration with Annette Lee at St. Cloud State University and Fond du Lac Tribal & Community College. Jim is currently a consultant with both NASA and NOAA using satellite visualization and storytelling, and he had an experiment on the last space shuttle STS-135. His Sisseton Dakota grandmother was a Red Day.

*William Wilson:* William Wilson (Lake Nipigon-Ojibwe) is from Ontario, Canada near Lake Nipigon (*Animbigon Zaaga'igan*-All You See Is Water). He was born and raised at his grandparent's house, speaking Ojibwe every day and living in a traditional way. Winter camp, snowshoeing, trapping, fishing, moose hunting and blueberry picking were a part of everyday life. William is a member of the Native Skywatchers team, a culture and language teacher, ceremonial/spiritual leader and a professional visual artist.

*Carl Gawboy:* Carl Gawboy (Boise Forte-Ojibwe) is from Ely, Minnesota, and is a prizewinning watercolorist. Recently, he has been a co-author with Ron Morton on the books, *Talking Rocks* and *Ancient Earth*. He retired from the College of St. Scholastica, where he taught in the Indian Studies Department.